\documentclass[a4paper,onecolumn,11pt,accepted=2021-03-31]{quantumarticle}
\pdfoutput=1
\usepackage[utf8]{inputenc}
\usepackage[english]{babel}
\usepackage[T1]{fontenc}
\usepackage{amsmath}
\usepackage{hyperref}
\usepackage{bm,bbold} 
\usepackage{tikz}
\usepackage{lipsum}
\usepackage[numbers,sort&compress]{natbib}

\usepackage{hyperref}
\hypersetup{
    colorlinks=true,       
    linkcolor=cyan,          
    citecolor=magenta,        
    filecolor=magenta,      
    urlcolor=cyan,           
    runcolor=cyan
}
\usepackage{qcircuit}

\newcommand{\bra}[1]{\langle #1 |}
\newcommand{\ket}[1]{| #1 \rangle}

\newcommand {\beq}{\begin{equation}}
\newcommand {\eeq}{\end{equation}}

\newcommand\e{{\textrm{e}}}

\newcommand{\bea}{\begin{eqnarray}}
\newcommand{\eea}{\end{eqnarray}}

\newcommand\tr{{\mbox{Tr\,}}}
\newcommand{\ignore}[1]{}

\begin{document}

\title{Quantum Algorithm for Simulating Hamiltonian Dynamics with an Off-diagonal Series Expansion}
\author{Amir Kalev}
\email{amirk@isi.edu}
\affiliation{Information Sciences Institute, University of Southern California, Arlington, VA 22203, USA}
\author{Itay Hen}
\email{itayhen@isi.edu}
\affiliation{Information Sciences Institute, University of Southern California, Marina del Rey, CA 90292, USA}
\affiliation{Department of Physics and Astronomy, and Center for Quantum Information Science \& Technology,University of Southern California, Los Angeles, California 90089, USA}
\maketitle

\begin{abstract}
\noindent We propose an efficient quantum algorithm for simulating the dynamics of general Hamiltonian systems. Our technique is based on a power series expansion of the time-evolution operator in its off-diagonal terms. The expansion decouples the dynamics due to the diagonal component of the Hamiltonian from the dynamics generated by its off-diagonal part, which we encode using the linear combination of unitaries technique. Our method has an optimal dependence on the desired precision and, as we illustrate, generally requires considerably fewer resources than the current state-of-the-art. 
We provide an analysis of resource costs for several sample models. 
\end{abstract}

\section{Introduction}
Simulating the dynamics of quantum many-body systems is a central challenge in Physics, Chemistry and the Material Sciences as well as in other areas of science and technology. 
While for classical algorithms this task is in general intractable, quantum circuits offer a way around the classical bottlenecks by way of `circuitizing' the time evolution of the system in question. However, present-day quantum computing devices allow for the programming of only small and noisy quantum circuits, a state of matters that places severe constraints on the types of applications these devices may be used for in practice. The qubit and gate costs of circuitization procedures have therefore rightfully become key factors in determining the feasibility of any potential application and increasingly more efficient algorithms are continuously being devised.  

We propose a novel approach to resource-efficient Hamiltonian dynamics simulations on quantum circuits that we argue offers certain advantages, which directly translate to a shorter algorithm runtime, over  state-of-the-art quantum simulation algorithms~\cite{Berry1,2018arXiv180500675H} (see Sec.~\ref{sec:comp} for a detailed comparison). We accomplish this by utilizing a series expansion of the quantum time-evolution operator in its off-diagonal elements wherein the operator is expanded around its diagonal component~\cite{ODE,ODE2,pmr}. This expansion allows one to effectively integrate out the diagonal component of the evolution, thereby reducing the overall gate and qubit complexities of the algorithm as compared to existing methods. 

In our approach, the time evolution is broken up into identical short-time segments, each of which is accurately approximated
using a number of terms in the off-diagonal series that is logarithmic in the inverse of the required precision. Each segment is then executed with the help of the linear combination of unitaries (LCU) lemma~\cite{Berry1}.
Our algorithm enables the simulation of a wide range of realistic models, including systems of spins, bosons or fermions.

The paper is organized as follows. In Sec.~\ref{sec:off}, we introduce the off-diagonal expansion insofar as it applies to the time-evolution operator. In Sec.~\ref{sec:qtea}, we present the Hamiltonian dynamics algorithm that we construct based on the expansion and in Sec.~\ref{sec:comp} we provide a comparison between the present scheme and two of the leading approaches to quantum simulations, the Taylor-series based approach of Berry {\it et al.}~\cite{Berry1} and the interaction-picture representation approach devised by Low and Wiebe~\cite{2018arXiv180500675H}. We examine several examples in some detail. A summary and some conclusions are given in Sec.~\ref{sec:summary}.

\section{Off-diagonal series expansion of the time-evolution operator\label{sec:off}}
We next derive an expansion of the time evolution operator based on the off-diagonal series expansion recently introduced in Refs.~\cite{ODE,ODE2,pmr} in the context of quantum Monte Carlo simulations.  
While we focus in what follows on time-independent Hamiltonians for simplicity, we note that an extension of the following derivation to include time-dependent Hamiltonians also exists~\cite{timeDepHamSim}.

\subsection{Permutation matrix representation of the Hamiltonian}
We begin by casting the Hamiltonian in the form
\beq \label{eq:basic}
H=\sum_{i=0}^M D_i P_i= D_0+ \sum_{i=1}^M D_i P_i \,,
\eeq
where the $D_i$ operators are diagonal in some known basis, which we will refer to as the computational basis and denote by $\{ |z\rangle \}$,  $P_0 :=\mathbb{1}$, and the $P_i$ operators (for $i>0$) are permutation operators, i.e.,  \hbox{$P_i| z \rangle=| z'(i,z) \rangle$} where $z'\neq z$,  i.e., they do not have any fixed points (equivalently, their diagonal elements are all zero). While the above formulation may appear restrictive it is important to note that any Hamiltonian can be written in this form. In particular, for models of spin-$1/2$ particles (qubits), the  $D_i$'s are diagonal in the Pauli-$Z$ basis, and the $P_i$'s are a tensor products of Pauli-$X$  operators,  $P_i \in \{\mathbb{1},X\}^{\otimes N}$ where $N$ is the number of spins.

We will refer to the principal diagonal matrix $D_0$ as the diagonal component of the Hamiltonian, while the set $\{D_i P_i \}_{i=1}^M$ of  off-diagonal operators (in the computational basis) give the system its  `off-diagonal dimension'.  We will call `diagonal energies' the (real) numbers obtained by acting with $D_0$ on computational basis states: \hbox{$D_0 | z \rangle = E_z| z \rangle$}. Similarly, by applying  the generalized permutation operator  $D_i P_i$ on a basis state, we obtain \hbox{$D_i P_i | z \rangle = d_i(z')| z' \rangle$}, where $d_i(z')$ will be in general a complex number ($z'$ depends on $z$ and $i$).  With these notations in hand, we move on to discuss the off-diagonal series expansion  of the time-evolution operator.

\subsection{Expansion of the time-evolution operator}
We next consider the evolution of a state under a time-independent Hamiltonian $H$ for time $t$. We expand the time evolution operator $\e^{-i H t}$ using the off-diagonal series expansion. 

We first consider the action of $\e^{-i H t}$ on a basis state $|z\rangle$:
\bea
\e^{-i H t} |z\rangle= \sum_{n=0}^{\infty}\frac{(-i t)^n}{n!} H^n | z \rangle =\sum_{n=0}^{\infty}\frac{(-i t)^n}{n!}  \Big(\sum_{i=0}^M D_i P_i\Big)^n | z \rangle =
 \sum_{n=0}^{\infty}   \frac{(-i t)^n}{n!}  \sum_{{S}_j^{(n)} \in \mathcal{S}_{n}} {S}_j^{(n)} | z \rangle \,,
\eea
where in the last step we have also expanded the multinomial $(\sum_{i} D_i P_i)^n$, and $\mathcal{S}_{n}$ denotes the set of all $(M+1)^n$ operators  that appear in the expansion of the multinomial  $(\sum_{i} D_i P_i)^n$. We proceed by `stripping' all the diagonal operators off the sequences ${S}_j^{(n)}$. We do so by evaluating their action on the relevant basis states, leaving only the off-diagonal operators unevaluated inside the sequence (for example, for the $n=2$ sequence $D_1P_1D_0 $ we write $D_1P_1D_0 \ket{z}=E_z D_1P_1\ket{z}=E_z D_1\ket{z_1}=E_{z}d_1(z_1)\ket{z_1}=E_{z}d_1(z_1)P_1\ket{z}$, where $\ket{z_1}=P_1\ket{z}$). Collecting all terms together, we arrive at: 
\bea\label{eq:snsq}
\e^{-i H t} |z\rangle=
 \sum_{q=0}^{\infty} \sum_{{\bf{i}}_q}  d_{{\bf i}_q} P_{{\bf{i}}_q} | z \rangle  \Bigl( \sum_{n=q}^{\infty} \frac{(-i t)^n}{n!}
 {\!\!\!\!\!\!\!\!}\sum_{\substack{k_0,\ldots,k_q \\\text{s.t.}\sum_i k_i=n-q} }{\!\!\!\!\!\!\!\!}{E^{k_0}_{z} \cdots E^{k_{q}}_{z_{q}}} \Bigr)\,,
\eea
where the boldfaced index ${\bf i}_q = (i_1,\ldots,i_q)$ is a tuple of indices $i_j$, with $j=1,\ldots, q$, each ranging from $1$ to $M$ and $P_{{\bf i}_q} := P_{i_q} \cdots P_{i_2}P_{i_1}$. In addition, similar to the diagonal energy $E_{z}=\langle z | D_0|z\rangle$, we denote  $E_{z_j}=\langle z_j | D_0|z_j\rangle$ are the energies of the states $|z\rangle,|z_1\rangle, \ldots, |z_q\rangle$ obtained from the action of the ordered $P_{i_j}$ operators appearing in the sequence $P_{{\bf i}_q}$ on $|z\rangle$, then on $|z_1\rangle$, and so forth. Explicitly, $P_{i_1}|z\rangle=|z_1\rangle, P_{i_2}|z_1\rangle=|z_2\rangle$, etc. (Note that  the sequence of states, and similarly the energies, should actually be denoted $|z_1(z,i_1)\rangle, |z_2(z,i_1,i_2)\rangle, \ldots$. For conciseness we will be using the abbreviations $|z_1\rangle,|z_2\rangle,\ldots$) Last, we have denoted $d_{{\bf i}_q}=\prod_{j=1}^q d_{i_j}(z_j)$  where 
\beq\label{eq:dj}
d_{i_j}(z_j) = \langle z_j | D_{i_j}|z_j\rangle
\eeq
can be considered  the `hopping strength' of $P_{i_j}$ with respect to $|z_j\rangle$ (see Ref.~\cite{ODE} for a complete and detailed derivation).

The infinite sum in parentheses in Eq.~(\ref{eq:snsq}) evaluates to the efficiently calculable \emph{divided-differences} representation~\cite{dd:67,deboor:05} 
\beq
 \sum_{n=q}^{\infty} \frac{(-i t)^n}{n!}
 {\!\!\!\!\!\!\!\!}\sum_{\substack{k_0,\ldots,k_q \\\text{s.t.}\sum_i k_i=n-q} }{\!\!\!\!\!\!\!\!}{E^{k_0}_{z} \cdots E^{k_{q}}_{z_{q}}} 
 = \e^{-i t [E_{z},\ldots,E_{z_q}]} \,,
\eeq
where the complex coefficient $\e^{-i t [E_{z},\ldots,E_{z_q}]}$ is the \emph{divided difference of the exponential function} over the multi-set of the energies $\{E_{z},\ldots, E_{z_q}\}$~\cite{dd:67,deboor:05} (more details can be found in Appendix~\ref{app:dd}). 

We may therefore write 
\beq
\e^{-i H t}|z\rangle  = V_z(t) \ket{z}\,,
\eeq 
where
\beq\label{eq:Vz}
V_z(t)=\sum_{q=0}^{\infty}  \sum_{{\bf i}_q}  \alpha_{{\bf i}_q}^{(z)}(t) P_{{\bf i}_q}
\eeq
and where we have denoted
\beq
\alpha_{{\bf i}_q}^{(z)}(t)  =\e^{-i  t [E_z,\ldots,E_{z_q}]} d_{{\bf i}_q}.
\eeq
(In the special case of $q=0$, $\alpha_{0}^{(z)}(t)=\e^{-i  t E_z}$.) In Appendix~\ref{app:ddDelta}, we show that one can pull out a global phase from $\e^{-i  t [E_z,\ldots,E_{z_q}]}$ to obtain $\e^{-i  t E_z} \e^{-i  t [\Delta E_z,\ldots,\Delta E_{z_q}]}$ where \hbox{$\Delta E_{z_j} = E_{z_j}-E_z$} (and specifically $\Delta E_{z} = 0$). Therefore, we can write $\alpha_{{\bf i}_q}^{(z)}(t)$ as:
\beq\label{eq:ddDelta}
\alpha_{{\bf i}_q}^{(z)}(t)  =\e^{-i  t E_z} \e^{-i  t [\Delta E_z,\ldots,\Delta E_{z_q}]} d_{{\bf i}_q} \,,
\eeq
where the divided-difference inputs are now energy differences rather than total diagonal energies.
\section{The Hamiltonian dynamics algorithm\label{sec:qtea}}
\subsection{Preliminaries}\label{sec:pre}
We first set some definitions and notations that will be used in the description of the algorithm. 
We denote the max norm of a matrix $A$ by \hbox{$\Vert A\Vert_{\rm max}=\max_{i,j}|A_{ij}|$}, where $A_{ij}$ are the matrix elements of $A$ in the computational basis. For every diagonal matrix $D_i$ (with $i>0$) we define the bounds $\Gamma_i \geq \Vert D_i\Vert_{\rm max}$, and  denote $\Gamma_{{\bf i}_q} = \prod_{j=1}^q \Gamma_{i_j}$. We define the dimensionless time $T= t\Gamma$ with $\Gamma=\sum_{i=1}^M \Gamma_i$,  the repetition number $r=\lceil T/\ln(2) \rceil$, and the short time interval $\Delta t = t/r \approx \ln(2) /\sum_{i=1}^M \Gamma_i$.  

\subsection{Decomposition to short-time evolutions}
To simulate the time evolution of $\e^{-i H t}$, we execute $r$ times in succession a short-time circuit for the operator 
\beq
U=\e^{-i H \Delta t} \,.
\eeq
Hereafter we omit the explicit dependence on $\Delta t$ for brevity. We write 
\begin{align}\label{eq:udt}
U&=U \sum_z |z\rangle \langle z|= \sum_z U|z\rangle \langle z|=\sum_z V_z|z\rangle \langle z|,
\end{align} 
where $V_z$ is given by Eq.~\eqref{eq:Vz} upon replacing $t$ with $\Delta t$.
We can rewrite $U$ as follows:
\begin{align}\label{eq:udt}
&U= \sum_z \e^{-i \Delta t E_z} \sum_{q=0}^{\infty}  \sum_{{\bf i}_q} \e^{-i  \Delta t [\Delta E_z,\ldots,\Delta E_{z_q}]} d_{{\bf i}_q}P_{{\bf i}_q} |z\rangle \langle z|
\nonumber\\
&= \Big(\sum_z \sum_{q=0}^{\infty}  \sum_{{\bf i}_q}  \e^{-i  \Delta t [\Delta E_z,\ldots,\Delta E_{z_q}]} d_{{\bf i}_q} P_{{\bf i}_q} |z\rangle \langle z|\Big) \e^{-i  \Delta t D_0} := U_{{\rm od}} \e^{-i  \Delta t D_0}\,.
\end{align} 
{We thus find that the off-diagonal expansion enables the effective decoupling of the evolution due to  the diagonal part of the Hamiltonian from the evolution due its off-diagonal part, allowing us $U$ as a product of   $U_{{\rm od}}$ and $\e^{-i  \Delta t D_0}$. In the special case where the off-diagonal part of the Hamiltonian is zero (thus, $d_{{\bf i}_q}=0$  for all ${{\bf i}_q}$), our method reduces directly to simulating diagonal Hamiltonians on a quantum computer.} The circuit implementation of the {diagonal} unitary $\e^{-i  \Delta t D_0}$ can be done with a gate cost ${\cal O}(C_{D_0})$ where $C_{D_0}$ is the gate cost of calculating a matrix element of $D_0$~\cite{NielsenChuang} (see Appendix~\ref{app:UD0} for more details). {This cost depends only of the locality of $D_0$, and is independent of its norm}. 

To simulate $U_{{\rm od}}$ we will use the LCU technique~\cite{Berry1}, starting with  writing $U_{{\rm od}}$ as a sum of unitary operators. To do that, we first note that \hbox{$\vert\e^{-i \Delta t [\Delta E_{z_q},\ldots,\Delta E_{z}]}\vert\leq \Delta t^q/ q!$} (this follows from the mean-value theorem for divided differences~\cite{deboor:05}). In addition, $d_{{\bf i}_q}/\Gamma_{{\bf i}_q}$ are complex numbers lying inside the unit circle.  Therefore, the norm of the complex number
\beq\label{eq:beta}
\beta_{{\bf i}_q}^{(z)}  =\frac{q!}{\Gamma_{{\bf i}_q}\Delta t^q} \e^{-i  \Delta t [\Delta E_{z},\ldots,\Delta E_{z_q}]} d_{{\bf i}_q}
\eeq
is not larger than 1. We can thus write $\beta_{{\bf i}_q}^{(z)}$ as the average of two phases
\begin{align}
&\beta_{{\bf i}_q}^{(z)} =\cos \phi_{{\bf i}_q}^{(z)} \e^{i \chi_{{\bf i}_q}^{(z)}}= \frac{1}{2}  \Big( \e^{i (\chi_{{\bf i}_q}^{(z)} +\phi_{{\bf i}_q}^{(z)})}+ \e^{i (\chi_{{\bf i}_q}^{(z)} -\phi_{{\bf i}_q}^{(z)})} \Big).
\end{align} 
Using this notation, we can write $U_{\rm od}$ as
\beq\label{eq:sum of unitary}
U_{\rm od}= \sum_{k=0,1} \sum_{q=0}^{\infty}  \sum_{{\bf i}_q} \frac{\Gamma_{{\bf i}_q} \Delta t^q}{2q!} U_{{\bf i}_q}^{(k)} \,,
\eeq
where 
\begin{align}
U_{{\bf i}_q}^{(k)}&= \sum_{z} \e^{i (\chi_{{\bf i}_q}^{(z)} +(-1)^k\phi_{{\bf i}_q}^{(z)})} P_{{\bf i}_q} |z\rangle \langle z|=P_{{\bf i}_q}\Phi_{{\bf i}_q}^{(k)} \,,
\end{align}
and  \hbox{$\Phi_{{\bf i}_q}^{(k)}=\sum_{z} \e^{i (\chi_{{\bf i}_q}^{(z)} +(-1)^k\phi_{{\bf i}_q}^{(z)})}  |z\rangle \langle z|$} is a (diagonal) unitary transformation. Since $P_{{\bf i}_q}$ is a bona-fide permutation matrix, it follows that $U_{{\bf i}_q}^{(k)}$  is a unitary transformation. Thus,  Eq.~\eqref{eq:sum of unitary} is  the short-time off-diagonal evolution operator $U_{\rm od}$ represented as a linear combination of unitary transformations. 

\subsection{The LCU setup}
To simulate the evolution under $U_{\rm od}$ on a finite-size circuit, we truncate the series,
 Eq.~\eqref{eq:sum of unitary}, at some maximal order $Q$, which leads to the approximate
\beq\label{eq:tildeU}
\widetilde{U}_{\rm od} =\sum_{k=0,1} \sum_{q=0}^{Q}  \sum_{{\bf i}_q} \frac{\Gamma_{{\bf i}_q} \Delta t^q}{2q!} U_{{\bf i}_q}^{(k)} \,.
\eeq
Since the coefficients of the off-diagonal operator expansion fall factorially with $q$ (similar to the truncation of the Taylor series in Ref.~\cite{Berry1}), setting 
\beq\label{eq:Q}
Q={\cal O}\Bigl(\frac{\log (T/\epsilon)}{\log \log (T/\epsilon)}\Bigr) \,,
\eeq
ensures\footnote{Formally, Eq.~\eqref{eq:Q} should read $Q={\cal O}\Bigl(\frac{\log (T/\epsilon)}{W( \log (T/\epsilon))}\Bigr)$ where $W(x)$ is the $W$-Lambert function~\cite{Corless1996}. The $W$-Lambert function can be approximated as $W(x)=\log x - \log \log x +o(1)$. } that the error per evolution segment is smaller than $\epsilon/r$:
\beq
\sum_{q=Q+1}^\infty\frac1{q!}\Bigr(\frac{T}{r}\Bigr)^q=\sum_{q=Q+1}^\infty\frac{\ln(2)^q}{q!}\leq\frac{\epsilon}{r} \,,
\eeq
{where the last step follows from the inequality $q!\geq(q/e)^q$}.
This choice ensures that the overall error is bounded by $\epsilon$ (as measured by the spectral-norm of the difference between the approximation and the true dynamics).

We next provide the details of the circuit we implement to execute the LCU routine and the resource costs associated with it. 

\subsubsection{State preparation}
The first ingredient of the LCU is the preparation of the state 
\begin{align}\label{eq:psi0}
|\psi_0\rangle &=\frac1{\sqrt{s}}  \sum_{q=0}^{Q}  \sum_{{\bf i}_q}\sqrt{\Gamma_{{\bf i}_q}\frac{ \Delta t^q}{q!}} |{\bf i}_q\rangle\Bigl(\frac{\ket{0}+\ket{1}}{\sqrt{2}}\Bigr)\,
\end{align}
where  \hbox{$|{\bf i}_q\rangle=\ket{i_1}\cdots\ket{i_q}\ket{0}^{\otimes(Q-q)}$} is shorthand for $Q$ quantum registers, each of which has dimension $M$ (equivalently, a quantum register with \hbox{$\lceil Q \log (M+1)\rceil$} qubits). In addition, since $\sum_{{\bf i}_q} \Gamma_{{\bf i}_q} = (\sum_i \Gamma_i)^q$ 
\beq\label{eq:s2}
s=\sum_{q=0}^{Q} \frac{ \Delta t^q}{q!}\sum_{{\bf i}_q}\Gamma_{{\bf i}_q}=\sum_{q=0}^{Q}\frac{ (\sum_i \Gamma_i\Delta t)^q}{q!}\approx 2\,, 
\eeq 
by construction {[recall that $\sum_{i=1}^M \Gamma_i \Delta t  \approx \ln(2)$]}.

We construct $\ket{\psi_0}$ in two steps: Starting with the state $|0\rangle^{\otimes Q}$ we transform the first register to the normalized version of
\beq
\ket{0}+\sqrt{\sum_{q=1}^Q\frac{(\Gamma\Delta t)^q}{q!}} \ket{1}. 
\eeq 
where $\Gamma=\sum_{i=1}^M \Gamma_i$. Then the $\ket{0}$ state of the $q$-th register ($q=2,\ldots,Q$) is transformed to the normalized version of 
\beq
\sqrt{\frac{(\Gamma\Delta t)^{q-1}}{(q-1)!}}\ket{0}+\sqrt{\sum_{q'=q}^Q\frac{(\Gamma\Delta t)^{q'}}{q'!}} \ket{1}. 
\eeq 
conditioned on the $(q-1)$-th register being in the $\ket{1}$ state. The resulting state, up to normalization, is 
\beq
|0\rangle^{\otimes Q} \to \sum_{q=0}^{Q}  \sqrt{\frac{(\Gamma\Delta t)^q}{q!}} |1\rangle^{\otimes q}|0\rangle^{\otimes (Q-q)}.
\eeq
The gate cost of this step is ${\cal O}(Q)$. Next, we act on each of the registers with a unitary transformation that takes a $\ket{1}$ state to the normalized version of $\sum_{i=1}^{M}\sqrt{ \Gamma_i}\ket{i}$. Finally we apply a Hadamard transformation on the last (qubit) register, resulting in the state $|\psi_0\rangle$. The gate  cost of this step is ${\cal O}(M)$~\cite{1629135}.
Denoting the unitary transformation that takes $|0\rangle^{\otimes Q+1}$ to $|\psi_0\rangle$ by $B$, we find that the gate cost of $B$ is ${\cal O}(M Q)$~\cite{Berry1}. 

\subsubsection{Controlled-unitary transformation\label{sec:cut}}
The second  ingredient of the LCU routine is the construction of the controlled operation
\beq\label{eq:CU}
U_C|{\bf i}_q\rangle|k\rangle|z\rangle= |{\bf i}_q\rangle|k\rangle U_{{\bf i}_q}^{(k)} |z\rangle=|{\bf i}_q\rangle|k\rangle P_{{\bf i}_q}\Phi_{{\bf i}_q}^{(k)}|z\rangle\,,
\eeq
where $|k\rangle$ is a single qubit {ancillary} state in the computational basis. The number of ancilla qubits here is $\lceil Q \log (M+1)\rceil+1$. Equation~\eqref{eq:CU} indicates that $U_C$ can be carried out in two steps: a controlled-phase operation ($U_{C\Phi}$) followed by a controlled-permutation operation ($U_{CP}$). 

\begin{figure}[h!]
\begin{center}
\hspace{1em}\Qcircuit @C=1em @R=0.2em  @!R{ 
\ket{{\bf i}_q}  & & \ctrl{3}&\qw& \ctrl{3}&\qw  \\
\ket{k}  & & \ctrl{2}&\qw&\ctrl{2}&\qw\\
\ket{z} &  & \ctrl{1}&\qw&\ctrl{1}&\qw & &\Phi_{{\bf i}_q}^{(k)} \ket{z}   \\
\ket{0} &  & \gate{U_{\chi\phi}}&\gate{U_{\rm ph}}&\gate{U_{\chi\phi}^\dagger}&\qw 
}
\end{center}
\caption{A circuit description of the controlled phase $U_{C\Phi}$ in terms of $U_{\chi\phi}$ and $U_{\rm ph}$. }
\label{fig:UCphi}
\end{figure}
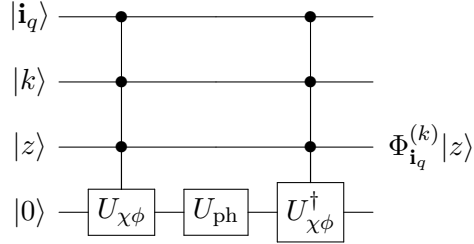

The controlled-phase operation $U_{C \Phi}$ requires a somewhat intricate calculation of non-trivial phases. We therefore carry out the required algebra with the help of additional ancillary registers and then `push' the results into phases. The latter step is done by employing the unitary 
\begin{align}
&U_{\text{ph}}|\varphi\rangle=\e^{-i \varphi}|\varphi\rangle \,,
\end{align}
whose implementation cost depends only on the precision with which we specify $\varphi$  and is independent of Hamiltonian parameters~\cite{NielsenChuang} (for completeness we provide an explicit construction of $U_{\text{ph}}$ in Appendix~\ref{app:Uph}).
With the help of the (controlled) unitary transformation
\beq\label{eq:F}
U_{\chi\phi}|{\bf i}_q\rangle|k\rangle|z\rangle\ket{0}= |{\bf i}_q\rangle|k\rangle|z\rangle|\chi_{{\bf i}_q}^{(z)} +(-1)^k\phi_{{\bf i}_q}^{(z)}\rangle \,,
\eeq
we can write
\beq
U_{C{\Phi}}=U_{\chi\phi}^\dagger (\mathbb{1} \otimes U_{\text{ph}}) U_{\chi\phi} \,,
\eeq
so that 
\beq\label{eq:ucphi}
U_{C{\Phi}}|{\bf i}_q\rangle|k\rangle|z\rangle=|{\bf i}_q\rangle|k\rangle\Phi_{{\bf i}_q}^{(k)}|z\rangle \,.
\eeq
This is illustrated in Fig.~\ref{fig:UCphi}.
Note that $U_{\chi\phi}$ is a `classical' calculation sending computational basis states to computational basis states. We provide an explicit construction of  $U_{\chi\phi}$ in Appendix~\ref{app:Phi}. 
We find that its gate and qubit costs are ${\cal O}(Q^2 +Q M (C_{\Delta D_0}+k_{\rm od} +\log M))$ and ${\cal O}(Q)$, respectively, where $C_{\Delta D_0}$ is the cost of calculating the change in diagonal energy due to the action of a permutation operator and  {$k_{\rm od}$ is an upper bound on the `off-diagonal locality', i.e., the locality of the $P_i$'s}~\cite{PhysRevA.52.3457,Berry1}. 

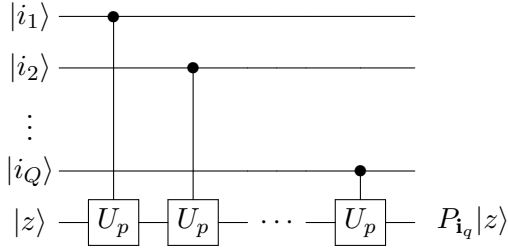
\begin{figure}[h!]
\begin{center}
\hspace{1em}\Qcircuit @C=1em @R=0.2em @!R { 
\ket{i_1}  & & \ctrl{4}&\qw&\qw&\qw &\qw&\qw&\qw &\\
\ket{i_2}  & & \qw&\ctrl{3}&\qw&\qw &\qw&\qw&\qw &\\
\vdots &  & & & & & & & \\
\ket{i_Q}  & & \qw&\qw&\qw&\qw &\qw&\ctrl{1}&\qw& \\
\ket{z} &  & \gate{U_p}&\gate{U_p}&\qw&\cdots & &\gate{U_p}&\qw& & P_{{\bf i}_q}\ket{z} 
}
\end{center}
\caption{A circuit description of $U_{C P}$.}
\label{fig:Cp}
\end{figure}

The construction of $U_{C P}$ is carried out by a repeated execution of the simpler unitary transformation \hbox{$U_p|i\rangle|z\rangle = |i\rangle P_i|z\rangle$}.  Recall that $P_i$ are the off-diagonal permutation operators that appear in the Hamiltonian. The gate cost of $U_p$ is therefore ${\cal O}(M (k_{\rm od} +\log M))$.
For spin models, each $P_i$ is a tensor product of up to $k_{\rm od}$ Pauli $X$ operators. Applying this transformation to the $Q$ ancilla quantum registers, we obtain \hbox{$|{\bf i}_q\rangle|z\rangle\to|{\bf i}_q\rangle P_{{\bf i}_q}|z\rangle$} with a gate cost of \hbox{${\cal O}(Q M (k_{\rm od} + \log M))$}. A sketch of the circuit is given in Fig.~\ref{fig:Cp}. We can thus conclude that the total gate cost of implementing $U_C$ is ${\cal O}(Q^2 +Q M (C_{\Delta D_0}+k_{\rm od} +\log M))$.  

\subsubsection{Oblivious amplitude amplification}
To realize $\widetilde{U}_{\textrm{od}}$, the LCU technique calls for the execution of a combination of the state preparation unitary $B$ and the controlled-unitary transformation $U_C$ which together form an oblivious amplitude amplification (OAA) procedure~\cite{Berry1}.

\begin{figure}[h!]
\begin{center}
\hspace{1cm}
\Qcircuit @C=0.7em @R=0.3em @!R{ 
\ket{\psi(t)}  &&&\ghost{\e^{-i\Delta t D_0}}&\qw&  \ghost{W}&\qw&\ghost{W^\dagger}&\qw&\ghost{-W} &\qw&&&&\ket{\psi(t+\Delta t)}\\
\ket{0} &&&\multigate{-1}{\e^{-i\Delta t D_0}}&\qw& \ghost{W}&\ghost{R}&\ghost{W^\dagger}&\ghost{R} &\ghost{-W}&\qw  \\
\ket{0}   &&&\qw&\qw& \multigate{-2}{W}&\multigate{-1}{R}&\multigate{-2}{W^\dagger}&\multigate{-1}{R} & \multigate{-2}{-W}&\qw 
}
\end{center}
\caption{A circuit diagram for a single short-time evolution step $U=\e^{-i H \Delta t}$. The bottom register consists of $Q$ sub-registers, each of which containing $\log M$ qubits. The middle line is a single-qubit register.} 
\label{fig:OAA}
\end{figure}
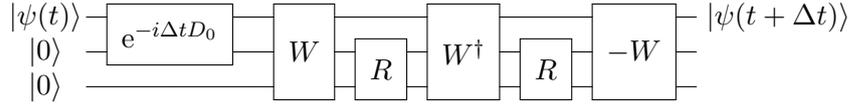

Let $\ket{\psi}$ be the current state of the system, then under the action of \hbox{$W=B^\dagger U_C B$},  the state becomes
\begin{align}
W\ket{0}^{\otimes Q+1}\ket{\psi}=\frac1{s}\ket{0}^{\otimes Q+1}\widetilde{U}_{\rm od}\ket{\psi}+\sqrt{1-\frac1{s^2}}\ket{\Psi^\perp},
\end{align}
such that $\ket{\Psi^\perp}$ is supported on a subspace orthogonal to $\ket{0}^{\otimes Q+1}$. If $s=2$ and $\widetilde{U}_{\rm od}$ is unitary then the OAA ensures that
\begin{align}\label{eq:A}
A\ket{0}^{\otimes Q+1}\ket{\psi}&=\ket{0}^{\otimes Q+1}\widetilde{U}_{\rm od}\ket{\psi},
\end{align}
where $A = -W R W^\dagger R W$ and $R=1-2 (\ket{0}\bra{0})^{\otimes Q+1}$. Under these conditions, the action of $W \, (\mathbb{1}\otimes\e^{-i\Delta t D_0})$ on the state at time $t$, namely $\ket{\psi(t)}$, advances it by one time step to $\ket{\psi(t+\Delta t)}$. This is illustrated in Fig.~\ref{fig:OAA}.

In Ref~\cite{Berry1}, a robust version of OAA was given for the case of non-unitary $\widetilde{U}_{\rm od}$ and $s \neq 2$. It is shown that if $\vert s - 2\vert = {\cal O}(\delta)$ and $\Vert \widetilde{U}_{\rm od}-U\Vert ={\cal O}(\delta)$, where $U$ is the (ideal) unitary transformation then
\beq
\Vert{\tr_{\rm anc}(PA\ket{0}^{\otimes Q+1}\ket{\psi})-U\ket{\psi}\bra{\psi}U^\dagger}\Vert={\cal O}(\delta)\,,
\eeq
where $\tr_{\rm anc}$ stands for trace over the ancilla registers {[recall that $s \approx 2$ as per Eq.~(\ref{eq:s2})]}.
Thus the overall error after $r$ repetitions is ${\cal O}(r\delta)$, so we require $\delta={\cal O}(\epsilon/r)$ to obtain an overall error of ${\cal O}(\epsilon)$. These conditions are satisfied with setting $\Delta t$ as in Sec.~\ref{sec:pre} and choosing $Q$ as in Eq.~\eqref{eq:Q}.

For convenience, we provide a glossary of symbols in Table~\ref{tbl:glossary}. A summary of the gate and qubit costs of the simulation circuit and the various sub-routines used to construct it is given in Table~\ref{tbl:resource}. \vspace{0.3cm}
\begin{table*}[t!]
\begin{center}
\begin{tabular}{ |c|l|}
  \hline
  Symbol&Meaning\\
  \hline
  $M$ & number of off-diagonal terms, c.f., Eq.~\eqref{eq:basic} \\
  $\Gamma_i$ & max-norm of $D_i$, $i=1,\ldots,M$  \\
 $T=t\sum_{i=1}^{M} \Gamma_i$ & dimensionless time  \\
  $Q$ & off-diagonal series expansion truncation order, $Q={\cal O}\Bigl(\frac{\log (T/\epsilon)}{\log \log (T/\epsilon)}\Bigr)$\\
  $k_{\rm d}$ & locality of $D_0$   \\
$k_{\rm od}$ & upper bound on locality of $P_i$  \\
$C_{D_0}$ & cost of calculating a diagonal energy (a single $D_0$ matrix element)   \\
$C_{\Delta D_0}$ & cost of calculating the change to a diagonal energy due to the \\
~&action of a $P_i$    \\
$C_{D}$ & cost of calculating a single $D_i$ matrix element ($i\neq0$)   \\
    \hline
 \end{tabular}
\end{center}
\caption{\label{tbl:glossary}{\bf Glossary of symbols.} }
\end{table*}
\begin{table*}[t!]
\begin{center}
\begin{tabular}{ |c||c|c|c| }
  \hline
  Unitary &Description&Gate cost&Qubit cost\\
  \hline \hline
 $\e^{-i\Delta t H}$ & short-time evolution&${\cal O}(C_{D_0}+Q^2 +Q M (C_{\Delta D_0}+k_{\rm od} +\log M))$ & ${\cal O}(Q\log M)$   \\\hline
$\e^{-i\Delta t D_0}$ & diagonal evolution &${\cal O}(C_{D_0})$ & ${\cal O}(1)$     \\\hline
$W$ & $W=B^{\dagger} U_C B$ &${\cal O}(Q^2 +Q M (C_{\Delta D_0}+k_{\rm od} +\log M))$& ${\cal O}(Q\log M)$    \\\hline
$B$ & LCU state preparation &${\cal O}(QM)$ & ${\cal O}(Q\log M)$    \\\hline
$U_C$ & LCU controlled unitary & ${\cal O}(Q^2 +Q M (C_{\Delta D_0}+k_{\rm od} +\log M))$ & ${\cal O}(Q\log M)$    \\\hline
$U_{CP}$ & controlled permutation &${\cal O}(QM(k_{\rm od}+\log M))$ & ${\cal O}(Q\log M)$    \\\hline
$U_{C\Phi}$ & controlled phase & ${\cal O}(Q^2 + Q M (C_{\Delta D_0}+k_{\rm od} +\log M))$ & ${\cal O}(Q\log M)$   \\\hline
\end{tabular}
\end{center}
\caption{\label{tbl:resource}{\bf A summary of resources for the circuit and the various sub-routines.} }
\end{table*}

\section{Comparison to existing approaches and examples\label{sec:comp}}

{In this section, we compare the resource costs  of our algorithm against those of two state-of-the-art existing approaches, and further provide a brief analysis of the complexity of our algorithm for a number of physical models.}

{Since our approach is based on an application of the LCU technique, we first compare  the resource costs of our algorithm's LCU sub-routine against that of the Taylor series-based method of Berry~{\it et al.}~\cite{Berry1}.} One of main differences in costs between the two series expansions stems from the different way in which the Hamiltonians are decomposed. In the Taylor series-based LCU the Hamiltonian is written as a sum of unitary operators $H=\sum_{i=1}^L c_i U_i$. For qubit Hamiltonians, these unitary operators will generally be tensor products of single-qubit Pauli operators (although of course in some cases, more compact decompositions can be found).  The off-diagonal decomposition, on the other hand, casts the Hamiltonian as a sum of generalized permutation operators, as given in Eq.~\eqref{eq:basic}; a representation that is generally considerably more compact. (For example, for qubit Hamiltonians, all operators that flip the same subset of qubits are grouped together.) This in turn implies that the number of terms in the decomposition of the Hamiltonian will generally be considerably smaller in the off-diagonal representation (i.e., $M \ll L$). This difference directly translates to reduced gate and qubit costs (a summary is given in Table~\ref{tbl:resource}). 

Another key difference is in the respective dimensionless time constants. In the off-diagonal expansion approach the dimensionless time constant is given by \hbox{$T=t \sum_{i=1}^M \Gamma_i$}, while in the Taylor series approach it is \hbox{$T'=t\sum_{i=1}^L c_i$}. In both approaches the dimensionless time determines the cutoff of the respective expansions, and controls the overall gate and qubit costs of the algorithm. Indeed, as we show below,  in general one has \hbox{$\sum_{i=1}^M \Gamma_i \ll \sum_{i=1}^L c_i$}, which directly translates to a reduced simulation cost in favor of the off-diagonal expansion.  To be more quantitative, we provide an explicit comparison between the off-diagonal and Taylor expansions for a few spin models in Table~\ref{tbl:comp}. The `price' we pay for the above savings is the additional ${\cal O}(Q^2)$ operations per time step required for calculating the divided-difference coefficient. {However, we note, that since $Q$ scales logarithmically with $T$, and $T$ is typically much smaller than $T'$, the advantages arising from the use of divided differences asymptotically outweigh this added complexity.}

\begin{table*}[t!]
\begin{small}
\begin{center}
\begin{tabular}{ |c|c|c| }
  \hline
  Hamiltonian&\multicolumn{2}{|c|}{$H=\sum_{ij} J_{ij} Z_i Z_j$}\\\hline
 Method & this paper &Taylor series LCU~\cite{Berry1}\\\hline
No. of LCU unitaries &0& $N^2$  \\\hline
 Dimensionless time ($T$)& 0& $t\sum_{ij} |J_{ij}|$ \\\hline
  Comments & $H$ is diagonal &- \\
  \hline
\end{tabular}
\\\vspace{0.3cm}
\begin{tabular}{ |c|c|c| }
  \hline
  Hamiltonian&\multicolumn{2}{|c|}{$H=\sum_{ij} J_{ij} Z_i Z_j +\sum_{ij} \tilde{J}_{ij} Z_i X_j  $} \\
  \hline
 Method & this paper &Taylor series LCU~\cite{Berry1}\\\hline
No. of LCU unitaries & $N+1$ & $2N^2$  \\\hline
 Dimensionless time ($T$)& $t{\sum_{j}|\sum_i\tilde{J}_{ij}|}$& $t\sum_{ij} (|J_{ij}|+|\tilde{J}_{ij}|)$ \\\hline
  Comments & $D_0=\sum_{ij} J_{ij} Z_i Z_j $ &- \\
~ &  $D_j=\sum_{i} J_{ij} Z_i$ &~ \\
  \hline
\end{tabular}
\\\vspace{0.3cm}
\begin{tabular}{ |c|c|c| }
  \hline
  Hamiltonian&\multicolumn{2}{|c|}{$H=\sum_{ijk} J_{ijk} Z_i Z_j Z_k+\sum_{ijk} \tilde{J}_{ijk} Z_i Z_j X_k $ }\\
  \hline
 Method & this paper &Taylor series LCU~\cite{Berry1}\\\hline
No. of LCU unitaries &$N$& $2N^3$  \\\hline
 Dimensionless time ($T$)& $t{\sum_k|\sum_{ij}\tilde{J}_{ijk}|}$ &  $t\sum_{ijk} (|J_{ijk}|+|\tilde{J}_{ijk}|)$ \\\hline
  Comments & $D_0=\sum_{ijk} J_{ijk} Z_i Z_j Z_k $&- \\
~ & $D_k=\sum_{ij} \tilde{J}_{ijk} Z_i Z_j$ &~ \\
  \hline
\end{tabular}
\end{center}
\end{small}
\caption{\label{tbl:comp}{\bf A comparison between the proposed method and the Taylor series-based approach~\cite{Berry1}.} In the table, $N$ denotes the number of qubits. {The table illustrates two important features of the proposed method as compared to the Taylor series-based approach for Hamiltonians written in the Pauli basis. Firstly, the Taylor series-based approach treats diagonal and off-diagonal components of the Hamiltonian in the LCU algorithm on an equal footing, while our method requires only the off-diagonal part as an input to the LCU algorithm. This is shown in the row labeled by `No. of LCU unitaries'. Secondly, in the Taylor series-based approach, each Pauli operator in the decomposition of $H$ is considered as a unitary, leading to a dimensionless time that is proportional to the sum of the absolute values of all the coefficients in the decomposition. In our approach on the other hand, all the diagonal operators that act in the same way on basis states are grouped into a single diagonal operator ($D_j$ in the table). Therefore, in our algorithm, the dimensionless time is proportional to the sum of the norm of all `grouped' diagonal operators (sans the diagonal component of the Hamiltonian). Due to this grouping,  the dimensionless time of the present method will be in general extensively smaller than that of the Taylor series-based method. Having a smaller dimensionless time translates to savings in gate and qubit resources as well as to a shorter runtime of the algorithm.}}
\end{table*} 

{As an alternative to the Taylor series-based algorithm, recently Low and Wiebe~\cite{2018arXiv180500675H} have proposed a framework within which the dynamics is formulated in the interaction picture using a (truncated) Dyson series expansion. There, the time-ordered multi-dimensional integrals of the Dyson series are  approximated via Riemann sums and implemented using control registers, ridding the simulation cost of most of its dependence on the diagonal component of the Hamiltonian. Our algorithm is similar in this way to the interaction picture approach, as the off-diagonal series expansion may be viewed as explicitly integrating the Dyson integrals (the reader is referred to Refs.~\cite{2020arXiv201009888K,ODE2} for more details pertaining to the relation between the off-diagonal series expansion and the Dyson series). There are however a few notable differences between the two algorithms, that translate to differences in resource scaling. The main difference is that the cost of the interaction picture approach still has a poly-logarithmic dependence on the norm of the diagonal part of the Hamiltonian while in our method the dynamics due to the diagonal part of the Hamiltonian is completely decoupled from that of its off-diagonal part. This decoupling ensures that our algorithm has no dependence on the norm of the diagonal component of the Hamiltonian.  In addition, the poly-logarithmic dependence on the various problem parameters in the interaction picture algorithm is obtained under certain assumptions on the gate cost of implementing specific unitary oracles. The power of the logarithmic polynomial ($\gamma$ in Ref.~\cite{2018arXiv180500675H}) is left undetermined in general. As mentioned above in the context of the Taylor series-based algorithm, the price paid for the decoupling is an additional ${\cal O}(Q^2)$ operations per time step, with $Q$, the expansion order, scaling logarithmically with the algorithm's dimensionless time $T$, which does not depend on the diagonal norm of the Hamiltonian.}

In the next subsections, we briefly analyze the off-diagonal circuit complexity for three models of scientific interest: the (Fermi-)Hubbard model, that of electronic structure and the Schwinger model. 

\subsection{The Fermi-Hubbard model}
We first examine the asymptotic cost of implementing the Fermi-Hubbard model~\cite{Hubbard}, which serves as a model of high-temperature superconductors.
The  Fermi-Hubbard Hamiltonian is given by
\bea\label{eq:FH-H-intro}
 H =U \sum_{i=1}^{N}   a^{\dagger}_{i \uparrow} a_{i \uparrow} a^{\dagger}_{i \downarrow} a_{i \downarrow}  - t_{\textrm h} \sum_{\langle i j\rangle\sigma}   \left(a^{\dagger}_{i \sigma} a_{j \sigma} + a^{\dagger}_{j\sigma} a_{i \sigma}\right) \,,
\eea
describing $N$ electrons with spin $\sigma \in \{\uparrow,\downarrow\}$ hopping between neighboring sites on a $d$-dimensional hyper-cubic lattice whose adjacency matrix is given by $\langle i j\rangle$ with hopping strength $t_{\textrm h}$. In addition, the model has an on-site interaction term with strength $U$ between opposite-spin electrons occupying the same site. 

The Fermi-Hubbard model can be mapped to qubits in a number of different ways~\cite{jwt,BK02,Verstraete_2005,DK20}.
For concreteness, we consider the Jordan-Wigner transformation (JWT)~\cite{jwt} which maps the second-quantized operator  $a_{j \sigma}$ to an operator on $j$ qubits according to
\beq
a_{j \sigma} \to\left( \prod_{k=1}^{j-1} Z_{k \sigma} \right)\frac{X_{j \sigma}-i Y_{j \sigma}}{2}
\eeq
so that \hbox{$a^{\dagger}_{j \sigma} a_{j \sigma} = (\mathbb{1}+Z_{j \sigma})/2$}.
To write the Fermi-Hubbard Hamiltonian in the form of Eq.\eqref{eq:basic}, we rewrite the JWT as
\begin{align}
a_{j \sigma} \to\left( \prod_{k=1}^{j-1} Z_{k \sigma} \right)\frac{\mathbb{1}+Z_{j \sigma}}{2}X_{j \sigma}\,,
\quad 
a_{j \sigma}^\dagger \to\left( \prod_{k=1}^{j-1} Z_{k \sigma} \right)\frac{\mathbb{1}-Z_{j \sigma}}{2}X_{j \sigma}.
\end{align}
Applying the transformation to the Hamiltonian, Eq.~\eqref{eq:FH-H-intro}, we arrive at:
\beq\label{eq:fh_ham_off}
H= D_0 +\sum_{\langle i j \rangle\sigma} D_{ij\sigma} X_{i\sigma} X_{j\sigma} \,,
\eeq
where we have identified 
\begin{align}
D_0 = \frac{U}{4} \sum_{j=1}^N (\mathbb{1}+Z_{j \uparrow})(\mathbb{1}+Z_{j \downarrow}) \quad {\textrm{and}} 
\quad 
D_{ij\sigma} = -\frac1{2} t_{\textrm h} \prod_{k=i}^{j} Z_{k\sigma}\,.
\end{align}
The product structure of $D_{ij\sigma}$ implies that their max-norm is simply given by $t_{\textrm h}$ for all $i,j,\sigma$. The number of off-diagonal terms is $M=N d$.  Therefore the dimensionless time $T$ of the simulation algorithm is $T=t M t_{\textrm h}=t N d t_{\textrm h}$. For comparison, in the Taylor series decomposition, the number of terms in the Hamiltonian is $L=3N+2 M$, and the dimensionless time parameter is $T' \sim 3 N U t +2 T$.  Note that due to the independence of $T$ on the on-site repulsion strength $U$, the off-diagonal expansion algorithm offers a favorable scaling  
as compared to the Taylor series-based LCU in the Mott-insulating regime $U\gg t_{\textrm h}$.

\subsection{Hamiltonian simulation of electronic structure}
Another model of major practical relevance is the simulation of electronic structure in the framework of which the stationary properties of electrons interacting via Coulomb forces in an external potential are of interest. This problem was recently analyzed in detail in Ref.~\cite{PhysRevX.8.011044}, where a `plane wave dual basis Hamiltonian' formulation was proposed, which diagonalizes the potential operator leading to a Hamiltonian representation with ${\cal O}(N^2)$ second-quantized terms, where $N$ is the number of basis functions. 

Using JWT to map the model to qubits, one arrives at
\begin{align}
\label{eq:jw_ham}
H & =  \sum_{\substack{p, \sigma \\ \nu \neq 0}}\left(\frac{\pi}{\Omega \, k_\nu^2} - \frac{k_\nu^2}{4 \, N} + \frac{2\pi}{\Omega} \sum_{j}\zeta_j \frac{\cos\left[k_\nu \cdot \left(R_j-r_p\right)\right]}{k_\nu^2}\right) Z_{p,\sigma}\\
& + \frac{\pi}{2\,\Omega } \sum_{\substack{(p, \sigma) \neq (q, \sigma') \\ \nu \neq 0}} \frac{\cos \left[k_\nu \cdot r_{p-q}\right]}{k_\nu^2} Z_{p,\sigma} Z_{q,\sigma'}+ \sum_{\nu \neq 0} \left(\frac{k_\nu^2}{2}- \frac{\pi \, N}{\Omega \, k_\nu^2} \right) \mathbb{1}\nonumber\\
& + \frac{1}{4\, N} \sum_{\substack{p \neq q \\ \nu\neq0, \sigma}} k_\nu^2 \cos \left[k_\nu \cdot r_{q - p} \right] \left(X_{p,\sigma} Z_{p + 1,\sigma} \cdots Z_{q - 1,\sigma} X_{q,\sigma} + Y_{p,\sigma} Z_{p + 1,\sigma} \cdots Z_{q - 1,\sigma} Y_{q,\sigma} \right)\nonumber,
\end{align}
where, $R_j$ and $r_p$ denote nuclei and electron coordinates, respectively, $\zeta_j$ are nuclei charges and $k_{\nu}$ is a vector of the plane wave frequencies at the $\nu$-th harmonic of the computational cell in three dimensions whose volume we denote by $\Omega$ (see Ref.~\cite{PhysRevX.8.011044}).  

The permutation matrix representation dictates that we write the Hamiltonian above as 
\beq\label{eq:es_ham_off}
H= D_0 + \sum_{p \neq q, \sigma} D_{pq\sigma} X_{p\sigma} X_{q\sigma}
\eeq
where all the diagonal terms are grouped together to form 

\begin{align}
\label{eq:jw_ham_D0}
D_0 & =  \sum_{\substack{p, \sigma \\ \nu \neq 0}}\left(\frac{\pi}{\Omega \, k_\nu^2} - \frac{k_\nu^2}{4 \, N} + \frac{2\pi}{\Omega} \sum_{j}\zeta_j \frac{\cos\left[k_\nu \cdot \left(R_j-r_p\right)\right]}{k_\nu^2}\right) Z_{p,\sigma}\\ 
&+ \frac{\pi}{2\,\Omega } \sum_{\substack{(p, \sigma) \neq (q, \sigma') \nu \neq 0}} \frac{\cos \left[k_\nu \cdot r_{p-q}\right]}{k_\nu^2} Z_{p,\sigma} Z_{q,\sigma'}
+ \sum_{\nu \neq 0} \left(\frac{k_\nu^2}{2}- \frac{\pi \, N}{\Omega \, k_\nu^2} \right) \mathbb{1}\nonumber,
\end{align}

and are integrated out of the LCU. Off-diagonal ($p\neq q$) terms are also grouped as
\bea
D_{pq\sigma} = \frac{1}{4 N} \sum_{\nu\neq0} k_\nu^2 \cos \left[k_\nu \cdot r_{q - p} \right] \left( Z_{p + 1,\sigma} \cdots Z_{q - 1,\sigma} \right) \left( \mathbb{1}_{pq}+Z_{p\sigma} Z_{q \sigma}\right)\,.
\eea
We notice that in the off-diagonal representation, the Hamiltonian of Eq.~\eqref{eq:es_ham_off} has a structure similar to that of Eq.~\eqref{eq:fh_ham_off}, with $k_{\rm od}=2$. Similar to the Fermi-Hubbard model, each $D_{ij\sigma}$ has a product structure and their max-norm is simply given by $\frac{1}{2 N} \vert\sum_{\nu} k_\nu^2 \cos \left[k_\nu \cdot r_{q - p} \right] \vert$ for all $p,q,\sigma$. The number of terms in the off-diagonal part of the Hamiltonian in this representation is $M=2(N^2-N)$, and thus the dimensionless time $T$ of the simulation algorithm is 
\beq
T=t (N-1)  \sum_{p \neq q} \Bigl\vert\sum_{\nu\neq0} k_\nu^2 \cos \left[k_\nu \cdot r_{q - p} \right] \Bigr\vert.
\eeq
For comparison, in the Taylor series-based LCU approach the number of terms in the Hamiltonian is $L=2N+6(N^2-N)$, and the dimensionless time parameter is 
\begin{align}
T'= t\Biggl( \sum_{\substack{p \neq q \\ \nu \neq 0}} \Bigl(\frac{\pi}{\Omega } \frac{1}{k_\nu^2}+ \frac{1}{2\, N}  k_\nu^2\Bigr)  \Bigl\vert\cos \left[k_\nu \cdot r_{q - p} \right] \Bigr\vert+ 2 \sum_{\substack{p \\ \nu \neq 0}}\Bigl\vert\frac{\pi}{\Omega \, k_\nu^2} - \frac{k_\nu^2}{4 \, N} + \frac{2\pi}{\Omega} \sum_{j}\zeta_j \frac{\cos\left[k_\nu \cdot \left(R_j-r_p\right)\right]}{k_\nu^2} \Bigr\vert\Biggr).
\end{align}
In particular, the dimensionless parameter in the current scheme depends only on the magnitude of the two-electron interaction and can take values much smaller than $T'$ due to a `destructive interference' of the cosine terms evaluated at different values of $[k_\nu \cdot r_{q - p}]$. 

\subsection{The Schwinger model}

The Schwinger model~\cite{PhysRev.128.2425}  is an Abelian low-dimensional gauge theory describing two-dimensional (one spatial plus time) Euclidean quantum electrodynamics with a Dirac fermion. 
Despite being a simplified model, the theory exhibits rich properties, similar to those seen in more complex theories such as QCD (e.g., confinement and spontaneous symmetry breaking). 

The model can be converted to an equivalent spin model~\cite{PhysRevResearch.2.023015,sc1,sc2}
whose Hamiltonian is
\bea
H = \frac1{2 a^2 g^2}\sum_{i=1}^{N-1} (X_i X_{i+1} + Y_i Y_{i+1})+\frac{m}{a g^2} \sum_{i=1}^N (-1)^i Z_i
+\sum_{i=1}^{N-1} \left[\epsilon_0\mathbb{1}+\frac{1}{2}\sum_{j=1}^i \left( Z_j+(-\mathbb{1})^j\right) \right]^2\,,
\eea
where $\epsilon_0$ is a constant (that can be set to zero), $g, m$ and $a$ are the fermion-gauge field coupling, mass and lattice spacing, respectively and $N$ is the number of lattice sites. 

In permutation matrix representation, the Hamiltonian is written as \hbox{$H=D_0 + \sum_i D_i X_i X_{i+1}$}
 where the diagonal component $D_0$ is given by 
\beq
D_0=\frac{m}{a g^2} \sum_{i=1}^N (-1)^i Z_i+\sum_{i=1}^{N-1} \left[\epsilon_0\mathbb{1}+\frac{1}{2}\sum_{j=1}^i \left( Z_j+(-\mathbb{1})^j\right) \right]^2
\eeq
and \hbox{$D_i =1/(2 a^2 g^2)(\mathbb{1}-Z_i Z_{i+1})$}.

It follows then that the number of off-diagonal terms is $M=N$ and the off-diagonal dimensionless time is $T=t N/(2 a^2 g^2)$. For comparison, in the Taylor series-based LCU approach the number of terms $L$ to which the Hamiltonian is decomposed is proportional to $N^2$ due to the diagonal term, and the dimensionless time parameter $T'$ scales as {${\cal O} (t(N^2 + m N/(a g^2)+ N/(a^2 g^2)))$}.  We thus find that the off-diagonal formulation provides in this case a scaling advantage over a Taylor series-based approach.

\section{Summary and conclusions\label{sec:summary}}
We proposed a quantum algorithm for simulating the dynamics of general time-independent Hamiltonians. 
Our approach consisted of expanding the time evolution operator using an off-diagonal series; a parameter-free Trotter error-free method that was recently developed in the context of quantum Monte Carlo simulations~\cite{ODE,ODE2,pmr}. This expansion enabled us to simulate the time evolution of states under general Hamiltonians using alternating segments of diagonal and off-diagonal evolutions, with the latter implemented using the LCU technique~\cite{Berry1}.   

We argued that our scheme provides considerable savings in gate and qubit costs for certain classes of Hamiltonians, specifically Hamiltonians that are represented in a basis in which the diagonal component is dominant. In fact, we find that for optimal savings one should choose the basis of representation such that the norm of the off-diagonal component of the Hamiltonian is minimal.

In this work, we focused only on time-independent Hamiltonians. The algorithm can be extended to the time-dependent case by writing the time-evolution operator in a Dyson series and appropriately discretizing the Dyson time integrals~\cite{timeDepHamSim}. 

We believe that further improvements to our algorithm can likely be made. In Appendix~\ref{app:alt}, we provide a slightly modified representation of the Hamiltonian which simplifies, to an extent, the circuit construction, specifically the implementation of the `classical' calculation $U_{\chi \phi}$, which requires additional auxiliary ${\cal O}(Q)$ ancillas beyond those required by the LCU. It would not be unreasonable to assume that it is possible to encode all the classical calculation directly into phases, eliminating this extra cost. 

\section*{Acknowledgements}
We thank Eleanor Rieffel for useful discussions and Yi-Hsiang Chen for valuable comments. 
Work by AK (quantum algorithm development) was supported by the U.S. Department of Energy (DOE), Office of Science, Basic Energy Sciences (BES) under Award DE-SC0020280. Work by IH (off-diagonal series expansion and resource analysis) was supported by the U.S. Department of Energy, Office of Science, Office of Advanced Scientific Computing Research (ASCR) Quantum Computing Application Teams (QCATS) program, under field work proposal number ERKJ347. 
Part of this work was done while AK was at the Joint Center for Quantum Information and Computer Science (QuICS) at the
University of Maryland.
\bibliographystyle{unsrtnat}
\bibliography{refs}

\appendix
\section{Divided differences}

\subsection{Definition and relevant properties\label{app:dd}}
We provide below a brief summary of the concept of divided differences, which is a recursive division process. This method is typically encountered when calculating the coefficients in the interpolation polynomial in the Newton form.

The divided differences~\cite{dd:67,deboor:05} of a function $f(\cdot)$ is defined as
\beq\label{eq:divideddifference2}
f[x_0,\ldots,x_q] \equiv \sum_{j=0}^{q} \frac{f(x_j)}{\prod_{k \neq j}(x_j-x_k)}
\eeq
with respect to the list of real-valued input variables $[x_0,\ldots,x_q]$. The above expression is ill-defined if some of the inputs have repeated values, in which case one must resort to the use of limits. For instance, in the case where $x_0=x_1=\ldots=x_q=x$, the definition of divided differences reduces to: 
\beq
f[x_0,\ldots,x_q] = \frac{f^{(q)}(x)}{q!} \,,
\eeq 
where $f^{(n)}(\cdot)$ stands for the $n$-th derivative of $f(\cdot)$.
Divided differences can alternatively be defined via the recursion relations
\bea\label{eq:ddr}
f[x_i,\ldots,x_{i+j}] = \frac{f[x_{i+1},\ldots , x_{i+j}] - f[x_i,\ldots , x_{i+j-1}]}{x_{i+j}-x_i} \,,
\eea 
with $i\in\{0,\ldots,q-j\},\ j\in\{1,\ldots,q\}$ and the initial conditions
\beq\label{eq:divideddifference3}
f[x_i] = f(x_{i}), \qquad i \in \{ 0,\ldots,q \}  \quad \forall i \,.
\eeq
A function of divided differences can be defined in terms of its Taylor expansion. In the case where 
$f(x)=\e^{-i t x}$, we have
\beq
\e^{-i t [x_0,\ldots,x_q]} = \sum_{n=0}^{\infty} \frac{(-i t)^n [x_0,\ldots,x_q]^n}{n!} \ . 
\eeq 
Moreover, it is easy to verify that
\beq  \label{eq:ts}
[x_0,\ldots,x_q]^{q+n} = \Bigg\{ 
\begin{tabular}{ l c l }
   $0$ & \phantom{$0$} & $n<0$ \\
  $1$ & \phantom{$0$} &  $n=0$ \\
  $\sum_{\sum k_j = n} \prod _{j=0}^{q} x_j^{k_j}$& \phantom{$0$} &  $n>0$ \\
\end{tabular}
 \,.
\eeq
One may therefore write:
\bea
\e^{-i t[x_0,\ldots,x_q]} &=& \sum_{n=0}^{\infty} \frac{(-i t)^n [x_0,\ldots,x_q]^n}{n!} =
\sum_{n=q}^{\infty} \frac{(-i t)^n [x_0,\ldots,x_q]^n}{n!} =
\sum_{n=0}^{\infty} \frac{(-i t)^{q+n} [x_0,\ldots,x_q]^{q+n}}{(q+n)!}\nonumber \,.
\eea
The above expression can be further simplified to
\bea
\e^{-i t[x_0,\ldots,x_q]} &=& \sum_{n=q}^{\infty} \frac{(-i t)^{n}}{n!} \sum_{\sum k_j = n-q} \prod _{j=0}^{q} (x_j)^{k_j} \,,\nonumber\\
\eea
as was asserted in the main text.

\subsection{Proof of Eq.~\eqref{eq:ddDelta}}\label{app:ddDelta}
Given a list of inputs $x_0,\ldots,x_q$, we prove that 
\beq
\e^{- i t [x_0,\ldots,x_q]} =\e^{- i t x} \e^{- i t [\Delta_0,\ldots,\Delta_q]} \,,
\eeq
where $x$ is an arbitrary constant and $\Delta_j = x_j -x$. 
By definition~\cite{deboor:05}, 
\beq
\e^{- i t [x_0,\ldots,x_q]} = \sum_j \frac{\e^{-i t x_j}}{\prod_{ k \neq j} (x_j -x_k)} \,
\eeq
(assuming for now that all inputs are distinct). It follows then that
\begin{align}
\e^{- i t [x_0,\ldots,x_q]}= \sum_j \frac{\e^{-i t (\Delta_j+x)}}{\prod_{ k \neq j} (\Delta_j -\Delta_k)}  =\e^{-i t x} \sum_j \frac{\e^{-i t \Delta_j}}{\prod_{ k \neq j} (\Delta_j -\Delta_k)}=\e^{-i t x} \e^{- i t [\Delta_0,\ldots,\Delta_q]} \,.
\end{align}
This result holds for arbitrarily close inputs and can be easily generalized to the case where inputs have repeated values. 

\section{Circuit construction of $\e^{-i\Delta t D_0}$}\label{app:UD0}
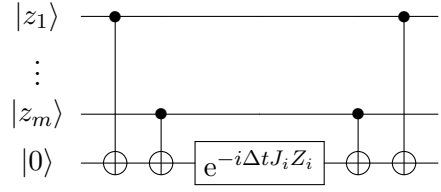
\begin{figure}[h!]
\begin{center}
\hspace{1em}\Qcircuit @C=0.75em @R=0.2em @!R { 
\ket{z_1}  & & &\ctrl{3}&\qw&\qw&\qw &\ctrl{3}&\qw\\
\vdots &  & & & & & & & & & \\
\ket{z_m}  & &  &\qw&\ctrl{1}&\qw &\ctrl{1}&\qw&\qw \\
\ket{0} &  & &\targ&\targ&\gate{\e^{-i\Delta t J_i Z_i}}&\targ &\targ &\qw
}
\end{center}
\caption{A circuit for $\e^{-i\Delta t J_i Z_i}$. The $m$ single-qubit registers control the application of  $\e^{-i\Delta t J_i Z_i}$ on the last single-qubit register.}
\label{fig:UD0}
\end{figure}

Assume that $D_0$ is a $k_{\rm d}$-local Hamiltonian on $n$ qubits, i.e., can be written as a sum of terms each of which acts on at most $k_{\rm d}$ qubits,
\beq
D_0=\sum_{i=1}^L J_i Z_i
\eeq
where $J_i\in\mathbb{R}$ and $Z_i$ is a shorthand for a specific tensor product of (at most) $k_{\rm d}$ single-qubit Pauli-$Z$ operators. Given this form of $D_0$ we can write
\beq
\e^{-i\Delta t D_0}=\prod_{i=1}^L \e^{-i\Delta t J_i Z_i}.
\eeq
Each unitary operator in the above product,  $\e^{-i\Delta t J_i Z_i}$, can be further simplified as
\beq
\e^{-i\Delta t J_i Z_i}=\sum_z\e^{-i\Delta t J_i (-1)^{\sum_{l=1}^m z_l} }\ket{z}\bra{z},
\eeq
where $\ket{z}$ is the computational basis state of the $m$ qubits on which $\e^{-i\Delta t J_i Z_i}$ acts ($m\leq k_{\rm d}$), i.e., \hbox{$z\in\{0,1\}^{m}$} and  $z_l=0,1$ is the $l$-th bit of $z$. Therefore, $\e^{-i\Delta t J_i Z_i}$ can be implemented using a single ancilla qubit and $2m$ CNOT gates~\cite{NielsenChuang}. A diagram is provided in Fig.~\ref{fig:UD0}.

\section{Circuit construction of $U_{\rm ph}$}\label{app:Uph}
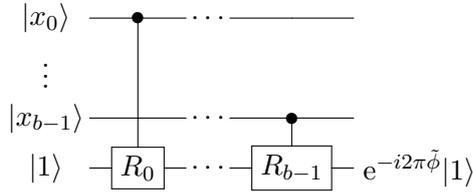
\begin{figure}[h!]
\begin{center}
\hspace{1em}\Qcircuit @C=0.75em @R=0.2em @!R { 
\ket{x_0}  & & &\ctrl{3}&\qw&\cdots& &\qw&\qw\\
\vdots &  & & & &  \\
\ket{x_{b-1}}  & &  &\qw&\qw&\cdots& &\ctrl{1}&\qw  \\
\ket{1} &  & &\gate{R_0}&\qw&\cdots& &\gate{R_{b-1}}&\qw &&&\e^{-i 2\pi \tilde{\phi}}|1\rangle
}
\end{center}
\caption{A circuit for $U_{\rm ph}$. The $b$ single-qubit registers control the application of  $R_j$ on the last single-qubit register.}
\label{fig:Uph}
\end{figure}

Following~\cite{NielsenChuang}, we construct a unitary \hbox{$U_{\text{ph}}|\tilde{\phi}\rangle=\e^{-i 2\pi \tilde{\phi}}|\tilde{\phi}\rangle$}, where \hbox{$\tilde{\phi}=\sum_{j=0}^{b-1}  2^j x_j$} is an approximate $b$-bit representation of \hbox{$\lceil 2^b\frac{\phi}{2\pi}\rceil$} for $\phi\in[0,2\pi)$.  We define a single qubit rotation
\beq
R_j=\begin{pmatrix}
    1 &0  \\
    0 & \e^{\frac{-2\pi i }{2^{b-j}}}  
\end{pmatrix}.
\eeq 
By applying $R_j$ on an additional single qubit in the state $\ket{1}$, controlled by the $j$-th register of $|\tilde{\phi}\rangle=|x_0,x_1,\ldots,x_{b-1}\rangle$, we obtain
\begin{align}
&|x_0,x_1,\ldots,x_{b-1}\rangle R_0^{x_0}\cdots R_{b-1}^{x_{b-1}}\ket{1}\\\nonumber&=|x_0,x_1,\ldots,x_{b-1}\rangle\e^{-i 2\pi \tilde{\phi}}\ket{1}=U_{\text{ph}}|\tilde{\phi}\rangle\ket{1}\,.
\end{align}
This is illustrated in Fig.~\ref{fig:Uph}.

\section{Implementation of  $U_{\chi\phi}$}\label{app:Phi}

The controlled unitary $U_{\chi\phi}$ essentially carries out a classical computation (it is a pure permutation, sending diagonal elements to diagonal elements). As such, its gate complexity can be given in terms of the classical cost of the calculation plus an incurred logarithmic overhead which comes from making the classical calculation reversible~\cite{rieffel:book}. 
The classical gate cost of calculating the complex number $\beta_{{\bf i}_q}^{(z)}$, Eq.~(\ref{eq:beta}), and therefore also that of \hbox{$\chi_{{\bf i}_q}^{(z)} +(-1)^k\phi_{{\bf i}_q}^{(z)}$} (recall that $\beta_{{\bf i}_q}^{(z)} =\cos \phi_{{\bf i}_q}^{(z)} \e^{i \chi_{{\bf i}_q}^{(z)}}$), consists of calculating (i) the product of $q$ off-diagonal Hamiltonian matrix elements, namely  $d_{{\bf i}_q}=\prod_{j=1}^q d_{i_j}(z_j)$, and (ii) the divided-differences of the exponential function with $q$ diagonal matrix elements as inputs (energy differences to be precise). The former can be calculated with ${\cal O}(1)$ registers and ${\cal O}(Q M\,C_{D})$ operations, where $C_{D}$ is the cost of calculating a single $D_j$ matrix element as this is a controlled operation on the evaluation of at most $Q$ off-diagonal matrix entries. The latter requires roughly ${\cal O}(Q M)$ operations to generate the energy differences in the worst case, and another ${\cal O}(Q^2)$ operations for calculating the divided difference given the inputs. The number of registers required for the above operation scales as ${\cal O}(Q)$~\cite{divDiffCalc}.  

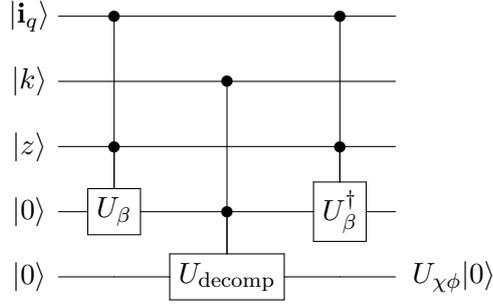
\begin{figure}[h!]
\begin{center}
\hspace{1em}\Qcircuit @C=1em @R=0.2em @!R { 
\ket{{\bf i}_q}  && \ctrl{3}& \qw& \ctrl{3}& \qw  \\
\ket{k}  && \qw& \ctrl{3}& \qw& \qw\\
\ket{z} && \ctrl{1}& \qw& \ctrl{1}& \qw  \\
\ket{0} && \gate{U_{\beta}}&\ctrl{1}& \gate{U_{\beta}^\dagger}& \qw\\
\ket{0} & & \qw & \gate{U_{\rm decomp}}& \qw& \qw&&U_{\chi\phi}|0\rangle
}
\end{center}
\caption{A circuit description of $U_{\chi\phi}$ in terms of the unitaries $U_{\beta}$ and $U_{\rm decomp}$. }
\label{fig:Uchiphi}
\end{figure}

With the above stated, we next present an explicit algorithm for constructing $U_{\chi\phi}$ for completeness. To that aim, we use two transformations, $U_{\beta}$ and $U_{\text{decomp}}$, defined as follows: 
\beq\label{eq:Ubeta}
U_\beta |{\bf i}_q\rangle|z\rangle|0\rangle  = |{\bf i}_q\rangle|z\rangle|\beta_{{\bf i}_q}^{(z)} \rangle \,,
\eeq
with $\beta_{{\bf i}_q}^{(z)}$ given in Eq.~(\ref{eq:beta}), and
\beq\label{eq:Odecomp}
U_{\text{decomp}} |\beta\rangle|k\rangle|0\rangle  = |\beta\rangle|k\rangle|\chi +(-1)^k \phi\rangle \,,
\eeq 
which decomposes a complex number $\beta=\cos \phi \e^{i \chi}$ (here, $|\beta|<1$) to the two angles $\phi$ and $\chi$. The third register of Eq.~\eqref{eq:Ubeta}, and similarly the first register of Eq.~\eqref{eq:Odecomp}, store complex numbers, that is, they each consist of two registers for their real and imaginary parts (hereafter, complex numbers inside kets indicate complex-number registers).  The gate cost of $U_{\text{decomp}}$ is a function of bit-precision only. 
Combined, the two sub-routines above yield
$
U_{\chi\phi}=U_\beta^\dagger\,U_{\text{decomp}}\,U_\beta$. 
This is illustrated in Fig.~\ref{fig:Uchiphi}. 

For the construction of $U_\beta$, it will be useful to rewrite the complex number $\beta_{{\bf i}_q}^{(z)}$ as
\beq
\beta_{{\bf i}_q}^{(z)} = \e^{- i \Delta t \Delta E_{(z,\ldots,z_q)}} \prod_j r_{i_j}
\eeq
where we have defined the (complex-valued) `effective energy difference' $\Delta E_{(z,\ldots,z_q)}$~\cite{ODE} such that
\beq
\e^{-i \Delta t \Delta E_{(z,\ldots,z_q)}}= \frac{q!}{(- i \Delta t)^q} \e^{-i  t [\Delta E_z,\ldots,\Delta E_{z_q}]} \,.
\eeq
Note that $\e^{-i \Delta t \Delta E_{(z,\ldots,z_q)}}$ is a complex number lying inside the unit circle. In addition, $r_{i_j}= - i d_{i_j}(z_j)/\Gamma_{i_j}$ are (normalized) matrix elements of $D_i$ ($i\neq0$) and are trivial to compute at the cost of evaluating an off-diagonal matrix element of the Hamiltonian. 

Thus, the only nontrivial component of $U_\beta$ is the sub-routine
\beq
U_{\rm dd}|{\bf i}_q\rangle|z\rangle |0\rangle = |{\bf i}_q\rangle|z\rangle |\Delta E_{(z,\ldots,z_q)} \rangle \,,
\eeq
which computes the `effective energy difference' (which should be followed by a ${\cal O}(1)$-cost circuit 
\hbox{$|\Delta E_{(z,\ldots,z_q)}\rangle|0\rangle \to |\Delta E_{(z,\ldots,z_q)}\rangle|\e^{-i \Delta t \Delta E_{(z,\ldots,z_q)}}\rangle$}.)

We next discuss a classically efficient method for calculating $\Delta E_{(z,\ldots,z_q)}$ given the sequence of energy differences $\{\Delta E_{z},\Delta E_{z_1},\ldots,\Delta E_{z_k}\}$.  
For simplicity we will assume the energy values are sorted (the divided difference of a function is invariant under a permutation of its inputs).
To carry out the calculation, we will use the divided differences recursion relations, Eq.~(\ref{eq:ddr}), which we will rewrite in terms of `effective energy differences'~\cite{ODE}:
\bea
\frac{(-i \Delta t)^q}{q!}\e^{-i \Delta t \Delta E_{(z,\ldots,z_q)}} =
 \frac{(-i \Delta t)^{q-1}}{(q-1)!}
\frac{
\left(\e^{-i \Delta t \Delta E_{(z,\ldots,z_{q-1})}} - \e^{-i \Delta t \Delta E_{(z_1,\ldots,z_q)}}\right)
}{\Delta E_z-\Delta E_{z_q}} \,.
\eea
Isolating $\Delta E_{(z,\ldots,z_q)}$, we arrive at 
\beq \label{eq:ede}
\Delta E_{(z,\ldots,z_q)}  = \bar{E} -\frac1{i \Delta t}\ln \frac{2 q \sin \Delta t \Delta \bar{E}}{\Delta t (E_{z_q}-E_{z})} \ ,
\eeq
where 
\bea
\bar{E} = (\Delta E_{(z_1,\ldots,z_q)}+\Delta E_{(z,\ldots,z_{q-1})})/2 \quad \textrm{and} \quad
\Delta \bar{E} =(\Delta E_{(z_1,\ldots,z_q)} - \Delta E_{(z,\ldots,z_{q-1})})/2 \,.
\eea
Thus Eq.~\eqref{eq:ede} provides a recursion relation for the effective energy differences. The initial condition for the above recursion is simply $E_{(z_i)} = E_{z_i}$ (we will sometimes denote a state $z$ by $z_0$ for notational convenience). 
In the limit where all energies in a sequence $\{z_i,\ldots,z_j\}$ are equal, i.e., $z_i=\ldots=z_j=z'$, the above relation neatly becomes \hbox{$\Delta E_{(z_i,\ldots,z_j)}=\Delta E_{(z')}=\Delta E_{z'}$}. 

\begin{figure}[th]
\begin{center}
\includegraphics[width=0.75\textwidth]{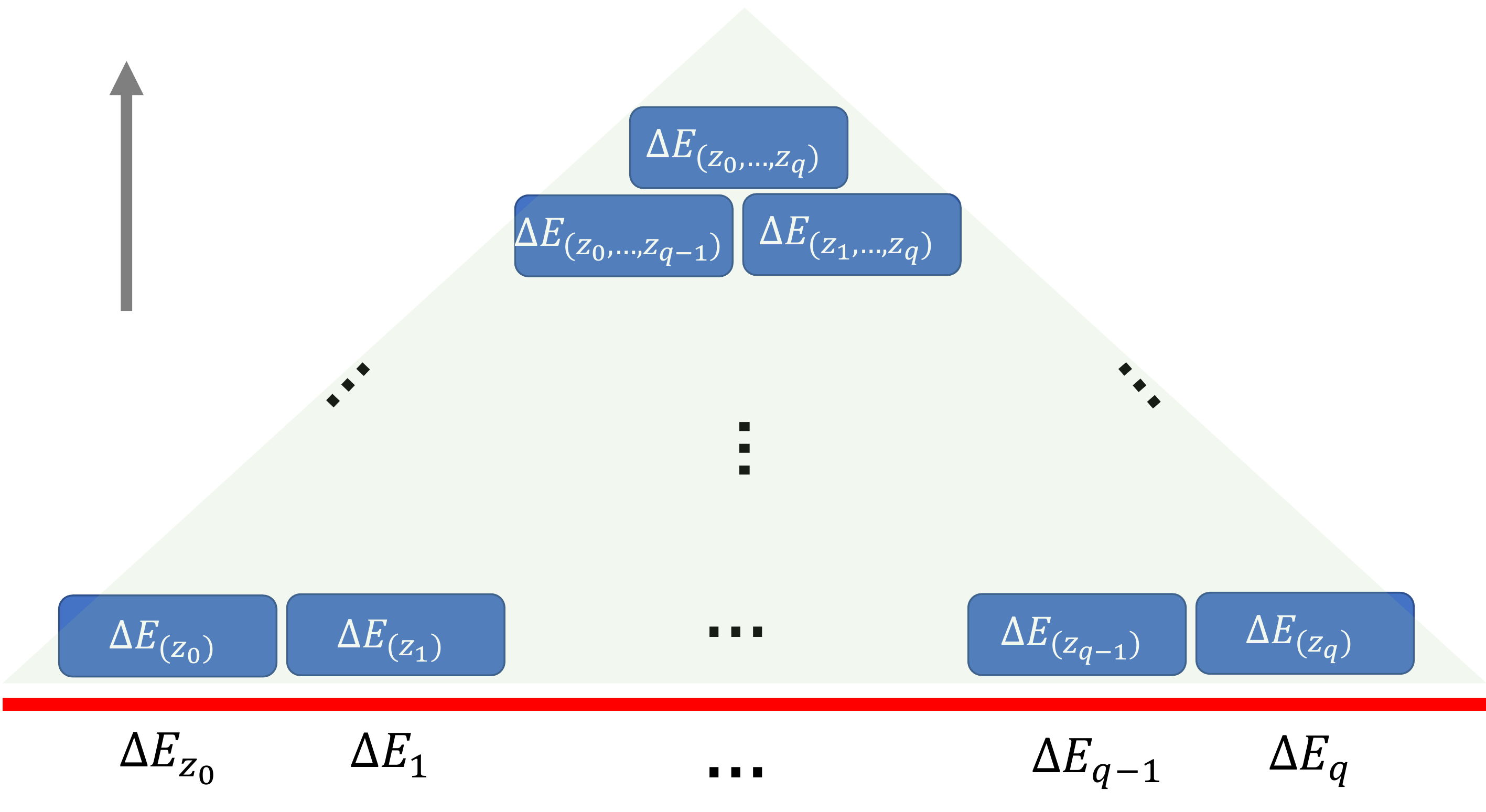}
\end{center}
\caption{Calculating the effective energy differences using a `pyramid' structure.   The evaluation of the divided differences of the exponential function of $q+1$ input energy differences consists of calculating each level of the pyramid starting at its base.  The values at the base of the pyramid $\Delta E_{(z_j)}$ are simply the energy inputs $\Delta E_{z_j}$ (shown as the red line at the bottom of the pyramid), with all identical energy differences placed together as a group (energies are assumed to be sorted).  To calculate the elements at the next level of the pyramid, we use the relation in Eq.~(\ref{eq:ede}).  This procedure is continued until the final level of the pyramid is evaluated, which corresponds to the desired effective energy difference $\Delta E_{(z,\ldots,z_q)}$.} 
\label{fig:pyr1}
\end{figure}

A convenient way to calculate the divided difference, equivalently the `effective energy difference' $\Delta E_{(z,\ldots,z_q)}$, relying on the recursion relations given above is using a `pyramid scheme' as illustrated in Fig.~\ref{fig:pyr1}. The base of the pyramid has $q+1$ elements, corresponding to the `initial' energies \hbox{$\Delta E_{(z_i)}=\Delta E_{z_i}$} with $i=0,\ldots, q$. Let us denote this as level zero.  Level one of the pyramid, which has $q$ elements only, is now evaluated as follows.  For each element at level one, we invoke the recursion relation Eq.~\eqref{eq:ede} using the two elements below it (see Fig.~\ref{fig:pyr1}) at level zero.
To avoid ill-defined ratios, we order the energies at level zero such that repeated values are grouped together. 
In this case, the evaluation of $\Delta E_{(z_i,z_{i+1})}$ for $\Delta E_{(z_i)}=\Delta E_{(z_{i+1})}$ gives $\Delta E_{(z_i,z_{i+1})} = E_{(z_i)}$. Similarly, every level-two element is calculated using the two level-one elements immediately below it. This procedure can be continued until the top level (level $q$) of the pyramid is reached, which gives the desired value of $\Delta E_{(z,\ldots, z_q)}$ the effective energy difference.
The above procedure requires ${\cal O}(Q)$ complex-valued registers and can be done reversibly with ${\cal O}(Q^2)$ operations. 

Because the routine just discussed for calculating divided differences requires the evaluation of logarithms and trigonometric functions, which are known to be somewhat cumbersome to implement on quantum computers~\cite{2018arXiv180512445H}, we also provide an alternative routine which is on the one hand slightly less efficient, requiring ${\cal O}(Q^2)$ registers but on the other hand uses only basic arithmetic operations.
To accomplish that, we first discuss an analogous classical method for calculating $\e^{-i \Delta t[\Delta E_{z_j},\ldots,\Delta E_{z_k}]}$ given a sequence of real numbers $\{\Delta E_{z_j},\ldots,\Delta E_{z_k}\}$. Our calculation will be based on the divided-differences Leibniz rule~\cite{dd:67,deboor:05} which states that for any value $\tau$
\beq\label{eq:leib}
\e^{-i \tau[\Delta E_{z_j},\ldots,\Delta E_{z_k}]} = \sum_{m=j}^k \e^{-i \frac{\tau}{2}[\Delta E_{z_j},\ldots,\Delta E_{z_m}]} \e^{-i \frac{\tau}{2}[\Delta E_{z_m},\ldots,\Delta E_{z_k}]} \,.
\eeq
We use this rule as a `time-doubling' mechanism for the exponent, from $\tau/2$ to $\tau$. 
Let us rephrase Eq.~\eqref{eq:leib} as 
\beq\label{eq:leib2}
e_{jk}(\tau)= 
\frac{1}{2^{k-j}}\sum_{m=j}^k {k-j \choose m- j}e_{jm}(\tau/2) e_{mk}(\tau/2)\,,
\eeq
where we have denoted 
\beq
e_{jk}(\tau)=\frac{(k-j)!}{(-i \tau)^{k-j}} \e^{-i \tau[\Delta E_{z_j},\ldots,\Delta E_{z_k}]}. 
\eeq
This certifies that the norms of the complex numbers we are dealing with are smaller than 1. 
For small enough values of $\tau$, that we denote by $\delta t$, 
we can write
\beq\label{eq:dd approx}
e_{jk}(\delta t) \approx \e^{-i \delta t \bar{\Delta}}=\prod_{m=j}^k \e^{-i \delta t\frac{\Delta E_{z_m}}{k-j+1}}  
\eeq
where $\bar{\Delta}=\sum_{m=j}^{k} \frac{\Delta E_{z_m}}{k-j+1}$. This approximation generates an error of
\beq
\left|e_{jk}(\delta t)- \e^{-i \delta t \bar{\Delta}} \right| < \delta t^2 \,
\eeq
(see Appendix~\ref{app:small} for proof).
We can therefore choose a large enough integer $\ell$ that is commensurate with the precision of the complex-valued registers and set  $\delta t= \Delta t/2^{\ell}$, and then repeatedly apply Leibniz's rule, Eq.~\eqref{eq:leib2}, $\ell$ times to calculate $e_{0q}(\Delta t)$. 

{With $\ket{\Delta E_z}\ket{\Delta E_{z_1}}\cdots\ket{\Delta E_{z_q}}$ as inputs, we then prepare the state:
\beq
|\xi (\delta t) \rangle_1|0\rangle_{2}\cdots|0\rangle_{\ell}
\eeq
where each of the $\ell$ registers above is a set of $Q^2/2$ complex-number registers and 
\beq
|\xi (\tau) \rangle =\bigotimes_{j=0}^Q\bigotimes_{k=j}^Q | e_{jk}(\tau)\rangle \,.
\eeq
Based on the approximation of Eq.~\eqref{eq:dd approx}, the state $|\xi (\delta t) \rangle$ can be prepared with cost ${\cal O}(Q^2)$. We populate the registers $|0\rangle_{2}\cdots|0\rangle_{\ell}$ in succession via the update rule
\beq
x_{jk}^{(\ell'+1)}\to x_{jk}^{(\ell'+1)} \oplus \frac{1}{2^{k-j}}\sum_{m=j}^{k} {k-j \choose m - j} x_{jm}^{(\ell')} x_{mk}^{(\ell')} \,,
\eeq
where $x_{jk}^{(\ell')}$ denotes the contents of the register in $(jk)$-th sub-register of the $\ell'$-th register (\hbox{$\ell'=1,\ldots, \ell-1$}).
The resultant state is
\beq
|\xi (\delta t) \rangle_{1}\cdots|\xi (2^{\ell} \delta t)\rangle_{\ell}=|\xi (\delta t) \rangle_{1}\cdots|\xi (\Delta t)\rangle_{\ell} \,.
\eeq
At this point we can uncompute registers $1, \ldots,\ell-1$ to free these registers arriving at the state
\beq
|0\rangle_1 |0\rangle_{2} \cdots |0\rangle_{\ell -1} |\xi (\Delta t)\rangle_{\ell}\,.
\eeq
The $(0q)$-th sub-register of the last register above contains the desired $\e^{-i \Delta t [E_z,\ldots,E_{z_q}]}$.}
{This circuit requires $Q(Q-1)/2 \times \ell$ complex registers. A more resource efficient version based on irreversible computation (and therefore also a similarly costly reversible version) of the circuit exists which only requires ${\cal O}(Q)$ registers and ${\cal O}(Q^2)$ gates~\cite{rieffel:book}. }

\begin{figure}[h!]
\begin{center}
\hspace{1em}\Qcircuit @C=0.75em @R=0.2em @!R { 
\ket{i_1}  & & &\ctrl{4}&\ctrl{4}&\qw&\qw&\qw&\qw&\cdots & &\qw&\qw&\qw&&\\
\ket{i_2}  & & &\qw&\qw&\qw&\ctrl{3}&\ctrl{4}&\qw&\cdots & &\qw&\qw&\qw&&\\
\vdots  \\
\ket{i_Q}  & & &\qw&\qw&\qw&\qw&\qw&\qw&\cdots & &\qw&\ctrl{5}&\qw&&\\
\ket{z} &  & &\gate{U_p}&\ctrl{1}&\qw&\gate{U_p}&\ctrl{2}&\qw&\cdots & &\qw&\ctrl{4}&\qw&&P_{{\bf i}_q}\ket{z}\\ 
\ket{0} &  & &\qw&\gate{U_{\Delta E}}&\multigate{1}{U_{+}}&\qw&\qw&\qw&\cdots & &\qw&\qw&\qw&&\ket{\Delta E_{z_1}}\\
\ket{0} &  & &\qw&\qw&\ghost{U_{+}}&\qw^{\hspace{0.75cm}\ket{\Delta E_{z_1}}}&\gate{U_{\Delta E}}&\qw&\cdots & &\qw&\qw&\qw&&\ket{\Delta E_{z_2}}\\
\vdots  \\
\ket{0} &  & &\qw&\qw&\qw&\qw&\qw&\qw&\cdots & &\qw&\gate{U_{\Delta E}}&\qw&&\ket{\Delta E_{z_Q}}
}
\end{center}
\caption{A circuit for $U_{\Delta}$, which calculates the energy differences $\Delta E_{z_i}$}
\label{fig:UDelta}
\end{figure}
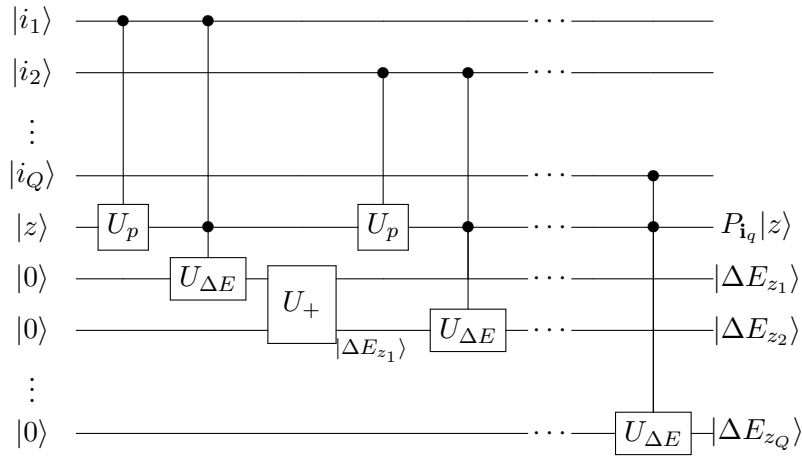

To complete our analysis, we provide next a circuit, $U_{\Delta}$, that generates the inputs to the divided-differences sub-routine, namely $\Delta E_{z_i}$. A sketch of the circuit is given in Fig.~\ref{fig:UDelta}. It requires $U_p$ (see Sec.~\ref{sec:cut}), the adder $U_{+}\ket{x}\ket{y}=\ket{x}\ket{x\oplus y}$
and  a sub-routine for calculating the difference in diagonal energy following a change in the input state $|z_{i-1}\rangle \to P_i|z_{i-1}\rangle=|z_i\rangle$, namely, 
\beq
U_{\Delta E} |i\rangle|z\rangle|y\rangle=|i\rangle|z\rangle|y+E_{z_{i}}-E_{z_{i-1}}\rangle \,.
\eeq
We note that, conveniently, the size of these energy-difference registers is not expected to grow with system size for physical systems.
The gate cost of $U_{\Delta E}$ is  ${\cal O}(M C_{\Delta D_0})$, where $C_{\Delta D_0}$ is the cost of calculating the change in diagonal energy due to the action of a single permutation operator, and therefore we can conclude that the gate cost of $U_{\Delta}$ is ${\cal O}(Q M (C_{\Delta D_0}+k_{\rm od} +\log M))$.

\section{Alternative description of the Hamiltonian\label{app:alt}}

A somewhat more efficient short-time evolution circuit can be obtained if we slightly modify the representation of the Hamiltonian. 
As before, every $D_i$ (for $i>0$) in $H$ can be written as $\Gamma_i (D_i/\Gamma_i)$ where $\Gamma_i$ is an upper bound on the norm of $D_i$ and the matrix elements of $(D_i/\Gamma_i)$ lie within the unit circle. Each such (diagonal) element can be written as a product  $\cos \theta \e^{i \chi}$. We can thus replace every $D_i$ with an average of two pure phase matrices with phases $\e^{i (\chi + \theta)}$ and  $\e^{i (\chi - \theta)}$ along their diagonals. This suggests the following representation of the Hamiltonian:
\bea
H= D_0 + \frac1{2} \sum_{i=1}^{M} \Gamma_i \left( \Theta^{(1)}_i + \Theta^{(2)}_i\right) P_i = D_0 + \sum_{i=1}^{M} \frac{\Gamma_i}{2} \left(P_i^{(1)} + P_i^{(2)} \right) \,.
\eea
where $\Theta^{(1)}_i$ and $\Theta^{(2)}_i$  are pure phase matrices, and $P_i^{(1/2)}=\Theta^{(1/2)}_i P_i$ are generalized permutations (and are of course unitary). 

Re-deriving the simulation algorithm using the above representation simplifies the `classical' calculation of the controlled phase $U_{C \Phi}$, Eq.~(\ref{eq:ucphi}), which now includes only the divided-difference calculation.

\section{Small-$\tau$ approximation of divided differences}\label{app:small}
{We show that 
\beq
 \left|\e^{-i \delta t[E_0,\ldots,E_q]} - \e^{-i \delta t \bar{E}} \right|  ={\cal O}(\delta t^2)\,,
\eeq
where 
\beq
\e^{-i  \delta t [E_0,\ldots,E_{q}]}= \frac{q!}{(- i \delta t)^q} \e^{-i  \delta t [E_0,\ldots,E_{q}]} \,,
\eeq
and
\beq
\e^{-i \delta t \bar{E}}=\prod_{m=j}^k \e^{-i \delta t \frac{E_m}{k-j+1}} \,.
\eeq
First, we observe that
\bea
\e^{-i  \delta t [E_0,\ldots,E_{q}]} &=& \cos \left(\delta t [E_0,\ldots,E_{q}]\right)-i \sin \left(\delta t [E_0,\ldots,E_{q}]\right)
\eea
and apply the mean-value theorem for divided differences~\cite{deboor:05}, which states that if $f(\cdot)$ is a real-valued function then
\beq
q! f[E_0,\ldots,E_q]=f^{(q)}(\tilde{E})
\eeq 
for some $\tilde{E}$ in the range $R_E=[\min_j E_j, \max_j E_j]$ and $f^{(q)}(\cdot)$ denotes the $q$-th derivative of $f(\cdot)$.}

{For small enough ranges $R_E$, $f(\cdot)$ is approximately linear, in which case $\tilde{E}$ will be the simple mean $\tilde{E} \approx \sum_j E_j/ (q+1)$. In this case, the error of the approximation will be of second order, meaning:
\beq
\left| q! f[E_0,\ldots,E_q] - f^{(q)}\left(\frac{\sum_j E_j}{q+1}\right)\right|  = {\cal O}(R_E^2)\,. 
\eeq
Choosing $f(\cdot)$ to be $\cos(\cdot)$ and $\sin(\cdot)$ with inputs $[\delta t E_0,\ldots, \delta t E_q]$, we arrive at:
\beq
\left| q! \e^{-i [\delta t  E_0,\ldots, \delta t  E_q]} - (-i)^q \e^{-i  \delta t \frac{\sum_j E_j}{q+1}} \right|  = {\cal O}(\delta t^2)\,. 
\eeq
Combining the above with the fact that
\beq
\e^{-i [\delta t  E_0,\ldots, \delta t  E_q]}  =\e^{-i \delta t [  E_0,\ldots, E_q]} / (\delta t)^q \,,
\eeq
completes the proof. }

\end{document}